\documentclass[12pt]{iopart}

\expandafter\let\csname equation*\endcsname\relax
\expandafter\let\csname endequation*\endcsname\relax

\usepackage{amsmath}
\usepackage{mathtools}
\usepackage{graphicx}  
\usepackage[normalem]{ulem}
\usepackage{xcolor}  
\begin{document}

\newcommand{\ud}[1]{{#1^{\dagger}}}
\newcommand{\au}{a_{\uparrow}}
\newcommand{\ad}{a_{\downarrow}}
\newcommand{\bra}[1]{\left\langle #1\right|}
\newcommand{\ket}[1]{\left| #1\right\rangle}
\newcommand{\braket}[2]{\langle #1|#2\rangle}
\newcommand{\ketbra}[2]{|#1\rangle\!\langle #2|}
\newcommand{\Imamoglu}{\u Imamo\=glu\xspace}
\newcommand{\Vuckovic}{Vu\u{c}kovi\'c}

\title[]{Two-photon correlations in detuned resonance fluorescence}

\author{Eduardo {Zubizarreta Casalengua}}
\address{Departamento de F\'isica Te\'orica de la Materia
  Condensada, Universidad Aut\'onoma de Madrid, 28049 Madrid,
  Spain}
\address{Faculty of Science and Engineering, University of
  Wolverhampton, Wulfruna St, Wolverhampton WV1 1LY, UK}

\author{Elena {del Valle}}
\address{Departamento de F\'isica Te\'orica de la Materia
  Condensada, Universidad Aut\'onoma de Madrid, 28049 Madrid,
  Spain}
\address{Institute for Advanced Study, Technische Universit\"at M\"unchen, 85748 Garching, Germany}

\author{Fabrice P. Laussy} 
\address{Faculty of Science and
  Engineering, University of Wolverhampton, Wulfruna St, Wolverhampton
  WV1 1LY, UK} 
\address{Russian Quantum Center, Novaya 100, 143025
  Skolkovo, Moscow Region, Russia}

\ead{fabrice.laussy@gmail.com} \vspace{10pt}
\begin{indented}
\item[]\today
\end{indented}

\begin{abstract}
  We discuss two-photon correlations from the side peaks that are
  formed when a two-level system emitter is driven coherently, with a
  detuning between the driving source and the emitter (quasi-resonance
  fluorescence). We do so in the context of the theories of
  frequency-resolved photon correlations and homodyning, showing that
  their combination leads to a neat picture compatible with
  perturbative two-photon scattering that was popular in the early
  days of quantum electrodynamics. This should help to control,
  enhance and open new regimes of multiphoton emission. We also
  propose a way to evidence the quantum coherent nature of the process
  from photoluminescence only, through the observation of a collapse
  of the symmetry of the lineshape accompanied by a surge of its
  intensity of emission. We discuss several of our results in the
  light of recent experimental works.
\end{abstract}

\section{Introduction: historical developments of resonance fluorescence}

Resonance fluorescence is the ``drosophila'' of quantum optics. It is
the simplest yet rich enough problem to capture many of the key
considerations on light-matter interaction, from quantization of the
light field up to the riddle of measurements and observations in
quantum mechanics. It consists of driving optically a two-level system
(e.g., an atomic transition, a spin, a semiconductor exciton, a
superconducting qubit, etc.)  with a coherent wave that has the same
or a close frequency than that of the spontaneous emission of the
emitter. The platform is both of great fundamental interest for the
understanding of basic aspects of quantum theory as well as from a
technological perspective for its prospects as a quantum emitter, not
only as a single-photon source but also in an unsuspected regime of
multiphoton emission.

The multiphoton problematic turns out to have been central to
theoretical modelling since the early days. As a basic problem of
light-matter interaction, the origin of resonance fluorescence goes
back to the dawn of quantum electrodynamics (which itself can be dated
with Dirac~\cite{dirac27a}). Pioneering contributions include those of
Weisskopf~\cite{weisskopf31a, weisskopf33a}, for scattering off the
ground and excited state of an atom, respectively.  A major and
recurrent work still of actuality, in the low-driving regime, is that
of Heitler~\cite{heitler_book54a} who reported his analysis directly
in the 3rd edition of his textbook ``the Quantum Theory of
Radiation'', in a chapter (absent in previous editions) titled
``resonance fluorescence'' (\S20), where he shows that the lineshape
of radiation is provided by the driving source itself as opposed to
the natural lineshape of the emitter. His analysis follows in essence
from the conservation of energy $\delta(\omega-\omega_\mathrm{L})$ of
the scattering process so that each photon from the source gets
scattered at the energy with which it impinged on the atom, whence the
result. The process is actually not as trivial as it looks, with
Heitler already observing that the radiation occurs ``as if two
independent processes, an absorption and a subsequent emission, took
place'', with the atom ``remembering'' (his term) ``before the
emission which quantum it has absorbed''. Seen in this way, it is less
obvious why the spontaneous emission character of the emitter plays no
role. It also brings forward that a two-photon process is involved in
this scattering. In fact, as we discuss further below, it turns out to
be of central importance for the photon statistics of resonance
fluorescence in this low-driving regime~\cite{lopezcarreno18b}: the
$\delta$-shaped scattered light itself is uncorrelated and becomes
antibunched only if also detecting the weak---but at the two-photon
level, essential---incoherent part of the spectrum, that is indeed
spread spectrally and originating from multiphoton events. This peak
is however very small in intensity as compared to the Rayleigh
peak. This led to some confusion in the
literature~\cite{nguyen11a,matthiesen12a} that we hope to have
clarified~\cite{hanschke20a} (see also~\cite{phillips20a}): although
the incoherent peak vanishes at low intensities in one-photon
observables, its contribution rules the photon-statistics (a two-photon
observable).

Multiphotons are even more prominent at higher driving. In a nonlinear
quantum-optics framework, several degenerate photons from the driving
source (which we shall from now on refer to as the ``laser'') can be
redistributed by the emitter at different energies so as to produce a
more complex spectral shape, known today to be a triplet with ratio of
peak heights 1:3:1 and with a splitting given by the laser intensity.
The problem was initially regarded as that of the competition between
spontaneous and stimulated emission, with a feeling shared by many
theorists of the time that spontaneous emission required quantization
of the field for a correct treatment.  The exact nature of this
spectral shape was the topic of some controvery, in particular it took
part in the debates initiated by Jaynes according to whom the light
field should not be quantized and his neoclassical theory (relying on
a nonlinear feedback from the radiation field back to the emitter)
should be used instead.  The neoclassicists were also experts in
solving the quantized version of a problem to provide what they
assumed were the wrong QED predictions, which is how, famously, the
Jaynes--Cummings model~\cite{jaynes63a} arose.  In this framework, the
quantized version of resonance fluorescence by Stroud~\cite{stroud71a}
(part of Jaynes' team) but at the one-photon level, led to incorrect
results, such as a 1:2:1 ratio of the peaks, in contrast to a
semiclassical treatment by Mollow~\cite{mollow69a} which was not
quantizing the light field but obtained, for the first time, the
correct lineshape. For this reason, this characteristic result, that
was originally referred to as the AC stark effect, became known as the
``Mollow triplet'' (it seems that Zoller~\cite{zoller78b} is the first
to have used this denomination).  Stroud \emph{et al.}  mention in
their conclusions that their analysis is ``incomplete in one important
aspect'': the truncation to one-photon emission. While they recognize
that ``in the real physical case there will be a cascade emitting many
spontaneous-emission photons'', they believed that antibunching would
make such successive emissions from a quantized model justifying their
approximation.  Further support for Stroud \emph{et al.}'s view came
from Smithers and Freedhoff~\cite{smithers74a} who claimed to have
included multiphoton effects and yet still arrive at the same
(incorrect) result as Stroud \emph{et al.}, but this was disputed by
Carmichael and Walls~\cite{carmichael75a} who observed that in their
treatment, ``Smithers and Freedhoff have not managed to include true
photon cascades but have simply followed a series of sequential
one-photon emissions''. Convincingly, by truncating their quantized
version to single-photon transitions, Carmichael and Walls showed how
they downgraded Mollow's spectrum to one with the same attributes as
Stroud's. Mollow's result, it must be emphasized, although not
quantizing the light-field, is not part of the neoclassical theory,
which treats spontaneous emission as a continuous process as opposed
to quantum jumps, leading to still further departures between the
various models. The reason why Mollow got the correct result is
interesting: ironically, it turns out that the semiclassical model is
equivalent to a multiphoton quantized model, and that multiphoton
effects are responsible for the lineshape, although this is a
single-photon observable. This has been recognized and commented by
various people at the time but the most insightful discussion seems to
be that of Mollow himself, in his 1975 follow-up
paper~\cite{mollow75a}. He carried on such a fully-quantum treatment,
including multiphoton contributions of all orders, and showed that the
$c$-number description of the laser does not spoil the fully-quantum
nature of the problem, as long as multiphoton effects are included.
These correspond to the back-reaction in Jaynes' neoclassical theory
and to what a modern treatment would qualify as virtual photons, i.e.,
the atom re-absorbing photons that it has just emitted. In this
context, the problem can be understood as a scattering one. In the
words of Mollow~\cite{mollow75a} `` the individual multiphoton
scattering processes [\dots] are concealed from view, with only their
accumulated effect exhibited''. We will come back to this important
observation later on.  Another key contribution to that approach of
resonance fluorescence comes from
Cohen--Tannoudji~\cite{cohentannoudji75a} and his
co-workers~\cite{cohentannoudji77a, cohentannoudji79a, dalibard83a,
  reynaud83a}, who provided both the so-called dressed-atom and a
perturbative scattering pictures. From this viewpoint, spontaneous
emission is not deemed central but is relegated to a secondary
plane. Instead, dressing the atom is considered first, yielding new
eigenstates for the system with an exact (all-order) treatment of the
light-matter coupling. Then spontaneous emission is brought back to
make the dressed atom cascade down its energy diagram and in this
process replacing photons from the laser by fluorescence photons from
the atom. This remains the most picturesque way to understand the
spectral shape of the Mollow triplet and can also account for a lot,
although not all, the phenomenology of correlations between the peaks.

\begin{figure}
  \centering
  \includegraphics[width=\linewidth]{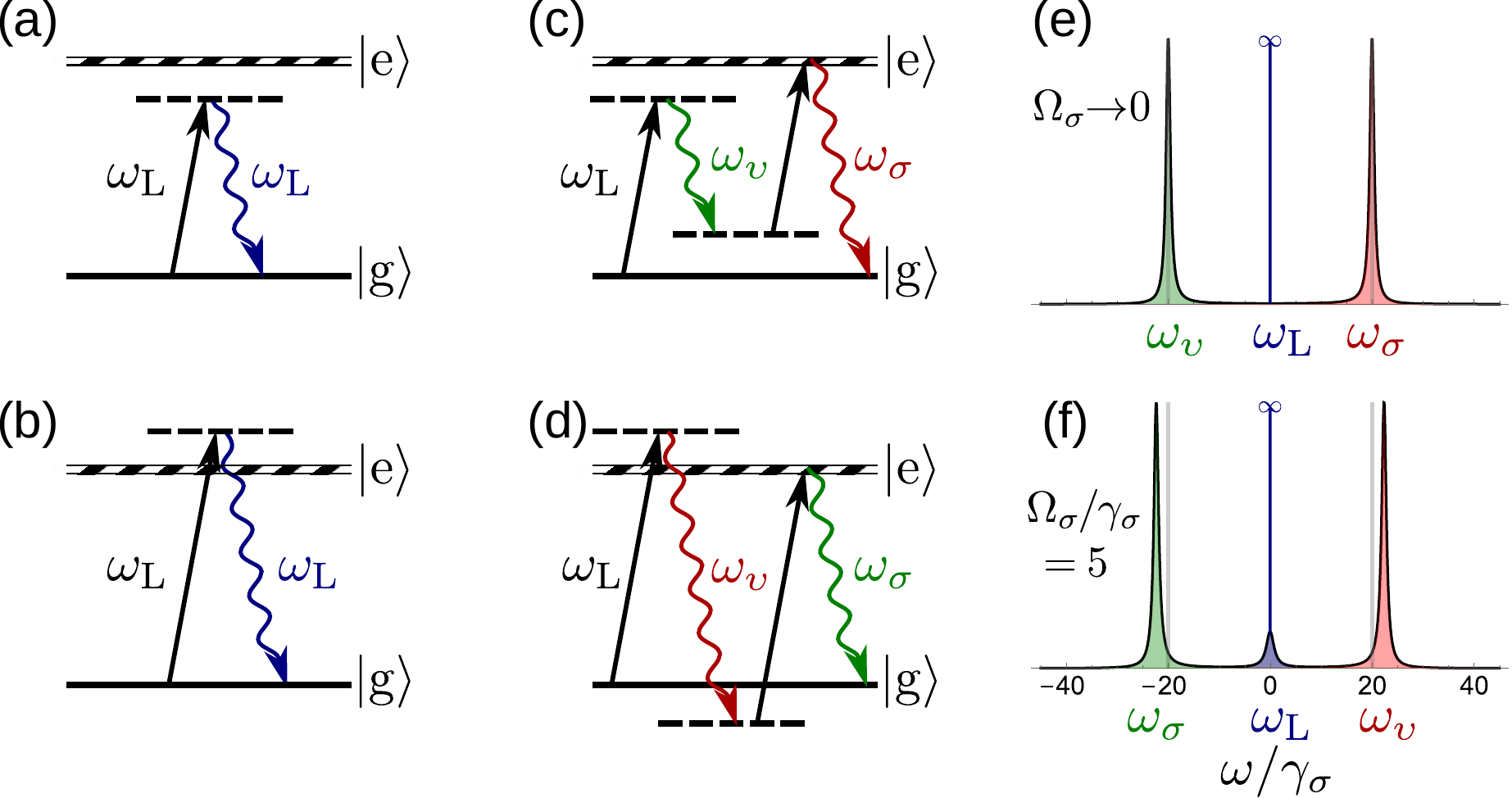}
  \caption{Scattering-picture of detuned resonance fluorescence, in
    the perturbative treatment of Dalibard and
    Reynaud~\cite{dalibard83a}. Regardless of detuning, the spectral
    shape is the same: the central peak is the Rayleigh~$\delta$ peak
    that elastically scatters the laser photon (depicted as a thin
    blue line topped by the~$\infty$ symbol). The detuning places the
    emitter on one side (upper row) or the other (lower row), while
    two-photon energy conservation places a copycat peak on the other
    side of the laser as result of the virtual states created to
    fill-in the loop (second column). The spectral shape is shown, in
    (e), for the Heitler regime of vanishing
    driving~$\Omega_\sigma\to0$ and, in~(f), at nonzero driving,
    producing an emerging fluorescence peak at the center of the
    triplet and slightly increasing the triplet splitting.}
  \label{fig:Sat24Sep165129BST2022}
\end{figure}

We now turn to the detuned resonance fluorescence, i.e., when the
driving laser is close to but not right at the energy of the two-level
system. In the earlier treatments, such as from Heitler, exact
resonance implied divergence and one of Heitler's inputs was precisely
to damp the system~\cite{heitler49a} so as to arrive to a physical
response for exact resonance. In a modern quantum-optical, master
equation approach, resonance is actually simpler while off-resonance
comes with additional subtleties but also with several advantages. Not
least is the fact that detuning helps the splitting of what always
remains a triplet. In fact, the first neatly resolved Mollow triplet
was out of resonance~\cite{schuda74a} and if one would stick to
resonance, it would then be apparently the improved setup of Walthers
that has reported the first resonant Mollow triplet, albeit in a
conference proceedings~\cite{walther75a} (the first report of an even
better triplet in a leading journal came however only a few months
later~\cite{wu75a}). Detuning also weakens the efficiency of the
coupling and what determines whether one is in the Heitler
(low-driving) or Mollow (high-driving) regime in this context is an
interesting question that we address
elsewhere~\cite{arXiv_zubizarretacasalengua22a}.  As was already
described by Mollow in his magnum opus~\cite{mollow69a}, when the 2LS
is detuned from the laser, one gets at low driving a doublet with a
peak centered on the atom and the other peak shifted by twice the
detuning, with the Rayleigh-scattered laser sitting in between, that
is to say, one always has a triplet in the non-detuned case. This is
shown in Fig.~\ref{fig:Sat24Sep165129BST2022}(e). Here it must be
appreciated that the two side peaks are vanishing
with~$\Omega_\sigma\to0$ as compared to the coherent peak in the
center, with a ratio
$8\Omega_\sigma^2/(\gamma_\sigma^2+4\Delta_\sigma^2))$ for their
respective intensities, i.e., most of the emission comes from the
central peak, just as the case of resonance where the Lorentzian
foothill is dwarfed by the Rayleigh peak. Here too, however,
two-photon observables, such as photon statistics, are ruled by the
interplay of the coherent and incoherent
emission~\cite{zubizarretacasalengua20b}, regardless of their relative
intensities. The only, but striking, difference is that this central
incoherent peak has now split in two.  With increasing driving, a
central incoherent peak grows at the laser position, becoming of
identical height with the side peaks
when~$\Omega_\sigma\approx\Delta_\sigma/\sqrt{2}$ (exactly so in the
limit~$\gamma_\sigma\ll\Delta_\sigma, \Omega_\sigma$) and for
higher-still driving converging towards a resonant Mollow triplet,
since detuning now becomes negligible as compared to driving. There is
therefore a smooth transition between the various
cases~\cite{arXiv_zubizarretacasalengua22a}.

A triplet structure comes with an obvious opportunity to correlate
photons from the various peaks. This was highlighted by
Cohen--Tannoudji and Reynaud~\cite{cohentannoudji79a} and implemented
by Aspect \emph{et al.}~\cite{aspect80a}. At low-enough driving but
with detuning to maintain a multi-peak structure, the perturbative
treatment is appealing and provides a remarkable and compelling
picture for the triplet, that is shown in
Figs.~\ref{fig:Sat24Sep165129BST2022}(a--d).  Here again, scatterings
between the states of the system gives the best phenomenological
explanation for the observed doublet (on the one hand) surrounding the
central peak (on the other hand).  This was discussed in details by
Dalibard and Reynaud~\cite{dalibard83a}. In addition to the Rayleigh
peak, that elastically scatters the photons and thus pins the peak at
the energy of the laser (1st column), the scheme also involves two
virtual states and a two-photon transition. One of the photons
originates from the excited-to-ground state transition with
energy~$\omega_\sigma$, but the other, not expected a priori,
originates from the virtual states created by the detuned laser which
needs them to close the two-photon emission process. This is at
energy~$\omega_\nu\equiv 2\omega_\mathrm{L}-\omega_\sigma$. Note that,
although the states are virtual, both photons are real. We will
clarify this statement in the following. The prediction from such a
picture, indeed confirmed experimentally~\cite{aspect80a}, is that
such a two-photon emission comes as a cascade: first the
virtual-state's photon then the emitter's.  While such considerations
have been made and confirmed a long time ago, recent reports by
Masters \emph{et al.}~\cite{arXiv_masters22a} of the observation of
bunching from the incoherent part of the spectrum of the low-driving
detuned resonance fluorescence, and by Long Ng \emph{et
  al.}~\cite{arXiv_longng22a} of cross-correlations from the
high-driving but also detuned Mollow triplet, bring back such
important questions in the limelight of modern setups and the improved
accuracy affordable today. In particular, modern authors envision
quantum-optical technological prospects, which were not at the core of
the preoccupations of the founding fathers who were worrying, instead,
on experimental validations of one or the other model of the theory of
light-matter interactions, from Dirac's quantum electrodynamics to
dissipative quantum optics. Such recent works can also participate to
the experimental validation of, or discrimination between, the more
refined theories available today, e.g., compare Figure~5 of
Ref~\cite{arXiv_longng22a} with Figure~4a of
Ref.~\cite{lopezcarreno18a} that itself includes, in addition to the
assumed exact cross-correlations between the peaks, results from
earlier works~\cite{schrama92a}.  More importantly, they also allow to
test more recent and new proposals regarding multiphoton emission from
resonance fluorescence. In the following, we discuss our own input to
this problem, which starts with the theory of frequency-resolved
photon correlations~\cite{delvalle12a}. We articulate our discussion
along the experimental findings of Long Ng, Masters \emph{et alii}.





\section{Frequency-resolved photon-correlation and homodyning}

Correlating the peaks in Fig.~\ref{fig:Sat24Sep165129BST2022} can be
done by filtering them first and directing the respective outputs to a
correlator (e.g., an Hanbury Brown--Twiss setup). This was already
discussed in precisely these terms by Cohen Tannoudji. Masters
\emph{et al.} chose a different strategy of filtering out the central
Rayleigh peak and auto-correlating the output. We will come back to
their approach but discussing first the more traditional one that has
been considered several times~\cite{aspect80a, ulhaq12a}, we must
highlight the input by Schrama \emph{et al.}~\cite{schrama92a}. They
implement the photodetection theory~\cite{nienhuis93a}, though at, or
close to, resonance and in the high (Mollow) driving regime, but more
importantly, with some approximations in the model to undertake the
complex calculations involved in the way the theory was then
formulated (as nested, time-ordered, high-dimensional integrals). We
provided an alternative, numerically exact as well as efficient,
formulation of the problem that allows us to compute faithfully
frequency-resolved $n$-photon correlations~\cite{delvalle12a}, in some
cases even analytically. This consists in enlarging the original
problem (in this case resonance fluorescence) with $n$ two-level
systems~$\varsigma_i$ (in this case, $n=2$) that are coupled both
through an Hamiltonian~$H_n$ and Liouvillian~$\mathcal{L}_n$ to the
system (in units of~$\hbar=1$):
\begin{equation}
  \label{eq:Thu29Sep164925CEST2022}
 \eqalign{\partial_t\rho=-i[\omega_\sigma\ud{\sigma}\sigma+\Omega(\sigma+\ud{\sigma}),\rho]+{\gamma_\sigma\over2}\mathcal{L}_\sigma\rho\cr
   {}-i\sum_{j=1}^2\big([\omega_i\ud{\varsigma_j}\varsigma_j,\rho]+\epsilon[\sigma\ud{\varsigma_j}+\ud{\sigma}\varsigma_j,\rho]\big)+\sum_{j=1}^2{\Gamma_j\over2}\mathcal{L}_{\varsigma_j}\rho
}
\end{equation}
where the first line in Eq.~(\ref{eq:Thu29Sep164925CEST2022}) is the
resonance fluorescence problem in its simplest possible and modern
quantum-optical formulation, while the second line implements the
sensor formalism where~$\omega_i$ set the frequencies which are to be
correlated while~$\Gamma_i$ set the filters' bandwidths.  One can then
(easily) solve this master equation without worrying for the
complicated frequency variables that otherwise enter at the level of
parameters in complex integrals, and which have been upraded here to
an operator. This makes the original evaluations in terms of folded
integrals turn to standard intensity--intensity correlations:
\begin{equation}
  \label{eq:Thu29Sep170946CEST2022}
  g_{\Gamma_1,\Gamma_2}(\omega_1,t_1,\omega_2,t_2)=\lim_{\epsilon\to0}{\langle{:}\ud{\varsigma_1}\varsigma_1(t_1)\ud{\varsigma_2}\varsigma_2(t_2){:}\rangle\over\langle\ud{\varsigma_1}\varsigma_1(t_1)\rangle\langle\ud{\varsigma_2}\varsigma_2(t_2)\rangle}\,,
\end{equation}
where the limit limit~$\epsilon\to 0$ is to be taken, providing finite
values for Glauber's correlators since both numerators and denominator
are of the same order.  Given that we will deal with steady-states
only, we will compute $g_{\Gamma_1,\Gamma_2}(\omega_1,\omega_2;\tau)$
with~$\tau\equiv t_2-t_1$ the time-delay between the photons, which
can be done with the quantum regression theorem. We need not elaborate
further here on the method itself, only emphasize again that it
provides the exact (according to the established theory of
photo-detection) correlations between the filtered photons, by
detectors with the respective bandwidths at the given time delay. Note
that if~$\omega_1=\omega_2$, this describes filtered
auto-correlations. In the following, we shall consider
that~$\Gamma_1=\Gamma_2$. We discuss the results in next Section. We
need however to first introduce another idea that relates to the
contribution of the Rayleigh peak.

The nature of antibunching in resonance fluorescence is very different
in the low-driving (Heitler) and high-driving (Mollow)
regimes~\cite{hanschke20a}. In the latter, it follows from the more
straightforward and popular picture of the two-level system being with
some probability in its excited state and, in its transition back to
the ground state, releasing a single photon. The density matrix in the
limit of infinite driving is
$\rho={1\over2}(\ketbra{0}{0}+\ketbra{1}{1})$. In stark contrast, in
the low-driving regime, antibunching arises from an interference
between the displaced squeezed thermal state in which the two-level
system is driven to leading-order, and the coherent state that is
imprinted by the driving laser~\cite{zubizarretacasalengua20b}. The
presence of squeezing as well as a strong Poisson content from the
laser means that the antibunched emission is of a more subtle
multiphoton character in this case. Photodetection tampers with the
squeezing in a way that is fundamentally equivalent to
frequency-filtering. This disturbs the interference that otherwise
yields a perfect antibuching. We have shown with L\'opez Carre{\~n}o
how one can, however, restore perfect antibunching by correcting for
the excess coherence from the laser as a result of filtering the tails
of the incoherent spectrum~\cite{lopezcarreno18b}. We later addressed
the case of detuned resonance fluorescence in the context of
filtered-homodyned correlations~\cite{lopezcarreno19a}, but we did so
for antibunching. In the next Section, we will instead phase-shift our
focus to consider the case of multiphoton emission, which is the one
revived by Masters \emph{et al.}~\cite{arXiv_masters22a} The formalism
is exactly the same, the regime and interpretations however deserve
considerations of their own. We also later generalized the scheme to a
large range of platforms where coherence is involved in some
form~\cite{zubizarretacasalengua20a}, in which case we have shown that
such multiphoton interferences are key to explain the structure and
type (conventional or unconventional) of a wide range of correlations,
from antibuching to superbunching. As a general statement, one can in
the framework that we have just laid down, seize additional control of
the system by homodyning, i.e., externally changing the nature
(constructive or destructive) of the interference and further
selecting auto or cross-correlations to characterize the system,
resulting in a much enhanced versatility and performance of its
emission. We do this in the following in the case of two-photon
emission in detuned resonance fluorescence.

\section{Removing the central peak}

\begin{figure}
  \centering
  \includegraphics[width=.5\linewidth]{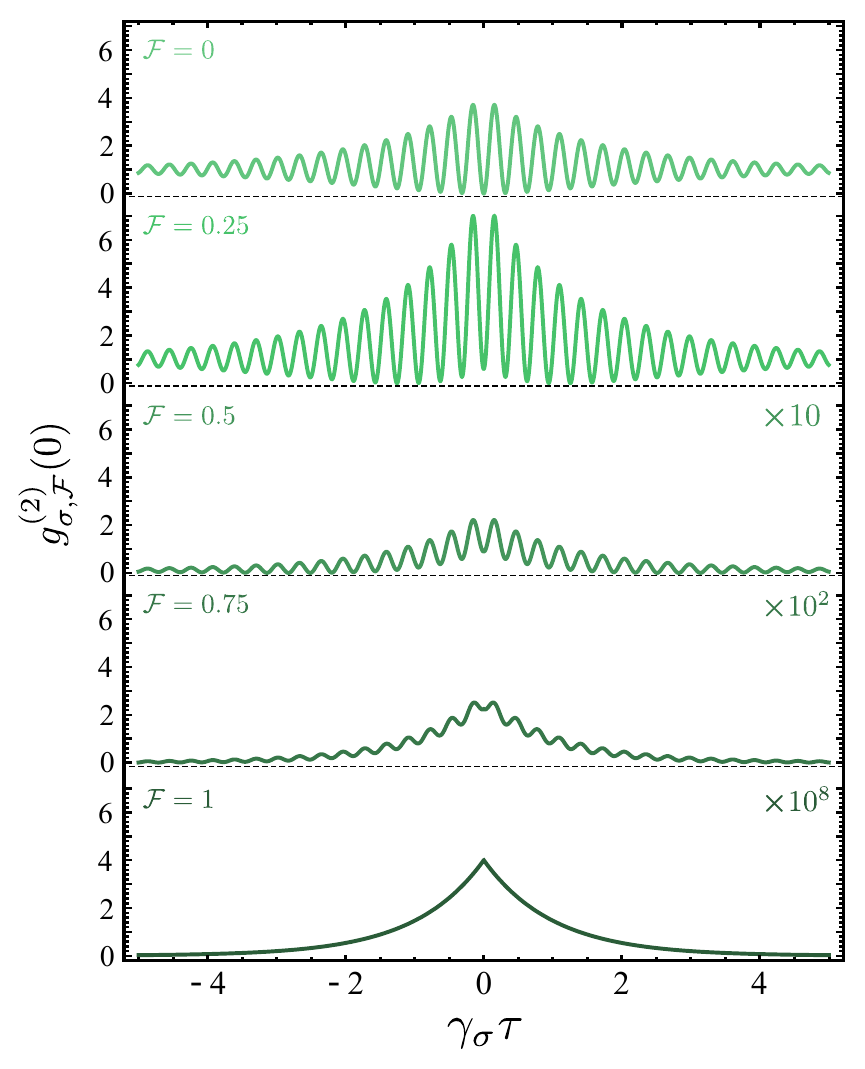}
  \caption{Two-photon correlation (no spectral filtering) as a
    function of the homodyning field that gradually suppresses the
    coherent (Rayleigh-scattered) central peak, from no
    supression~($\mathcal{F}=0$) to complete
    suppression~($\mathcal{F}=1$). The latter case is the ideal
    realization of Masters \emph{et al.}'s suppression of the coherent
    part. We find a good agreement with their result but with the more
    straightforward form of supressing the oscillations.}
  \label{fig:Tue27Sep194057CEST2022}
\end{figure}

We now turn to the exact computation of correlations in the case where
the laser is (with no loss of generality, since solutions are
symmetric) red-detuned at~$\omega_\mathrm{L}$ as compared to the atom
at~$\omega_\sigma=\omega_\mathrm{L}+\Delta_\sigma$
(cf.~Fig.~\ref{fig:Sat24Sep165129BST2022}(c,e)). The main question
addressed by Masters \emph{et al.} is whether the side peaks are
bunched, what they interpret as simultaneous two-photon emission. We
come back to qualify further this interpretation but first consider
the question itself. In their case, they used a narrowband notch
filter to attenuate the central peak, and let otherwise pass the other
photons, i.e., from the side peaks, which they autocorrelate. The
ideal version of this experiment is precisely our homodyning scheme,
unfiltered, where we remove the central peak by destructive
interference.  Indeed, we have checked that the two-photon coincidence
for the notch filter of width~$\Gamma$ and centered
at~$\omega_\sigma$ 
when filtering out the coherent Rayleigh peak, agrees in the limit
$\Omega_\sigma\to0$ with the full-homodyned zero-delay coincidence,
that is given by
$g^{(2)}_{\sigma,\mathcal{F}=1}(0)=(\gamma_\sigma^2+4\Delta_\sigma^2)(\gamma_\sigma^2+4[8\Omega_\sigma^2+\Delta_\sigma^2])/(64\Omega_\sigma^4)$.
The general and time-resolved case is shown in
Fig.~\ref{fig:Tue27Sep194057CEST2022} from, bottom, no homodyning
(corresponding to
$\mathcal{F}=0$), up to, top, full-destructive homodyning
($\mathcal{F}=1$) that completely removes the laser contribution.  In
the first case, we are looking there at the full autocorrelation of
the light itself. This is a well-known result: the antibunching is
perfect but oscillates strongly, which is usually (and correctly)
interpreted as the effect of detuning. As the coherent peak is
removed, one can see how antibunching is gradually lost, along with
the oscillations (although the two peaks remain detuned) to give rise
to bunching. This is the same result as obtained by Masters \emph{et
  al.} who, both in their experiment and theoretical treatments,
report a more complicated structure, due to the filtering. For more
stringent filtering their result reverts in character but this needs
not concern us here as this involves even more complicated but
technical effects that are well described by the authors.  We believe
that our description of their idea provides its cleanest and ideal
formulation.  Now turning to its interpretation, one must note that
the interference is perfect between the two splitted incoherent peaks
and the central coherent peak: regardless of the detuning, detecting
all frequencies yield perfect antibunching. This must be kept in mind
when considering the interpretation of the perturbative picture of
Fig.~\ref{fig:Sat24Sep165129BST2022}.  The coherent peak, that plays
no role in the diagram itself, is fundamental to decide of the outcome
of the actual detection. A possible interpretation is that the
coherent peak is responsible for the synchronization of the two
scattered photons from the diagram, so that they do not appear
together in the emission but are modulated in time, so as to ensure
perfect antibunching
at~$\tau=0$ but resulting in strong-bunching shortly afterwards. That
is another, more sophisticated but equally valid, way to understand
the oscillations
in~$g^{(2)}(\tau)$. Still another understanding~\cite{lopezcarreno19a}
is that such a detection at all frequencies removes the
indistinguishability between the photons and one can thus no longer
assert that their detuning leads to a spectral distinction.

\section{Cross-correlations of the side peaks}

\begin{figure}
  \centering
  \includegraphics[width=\linewidth]{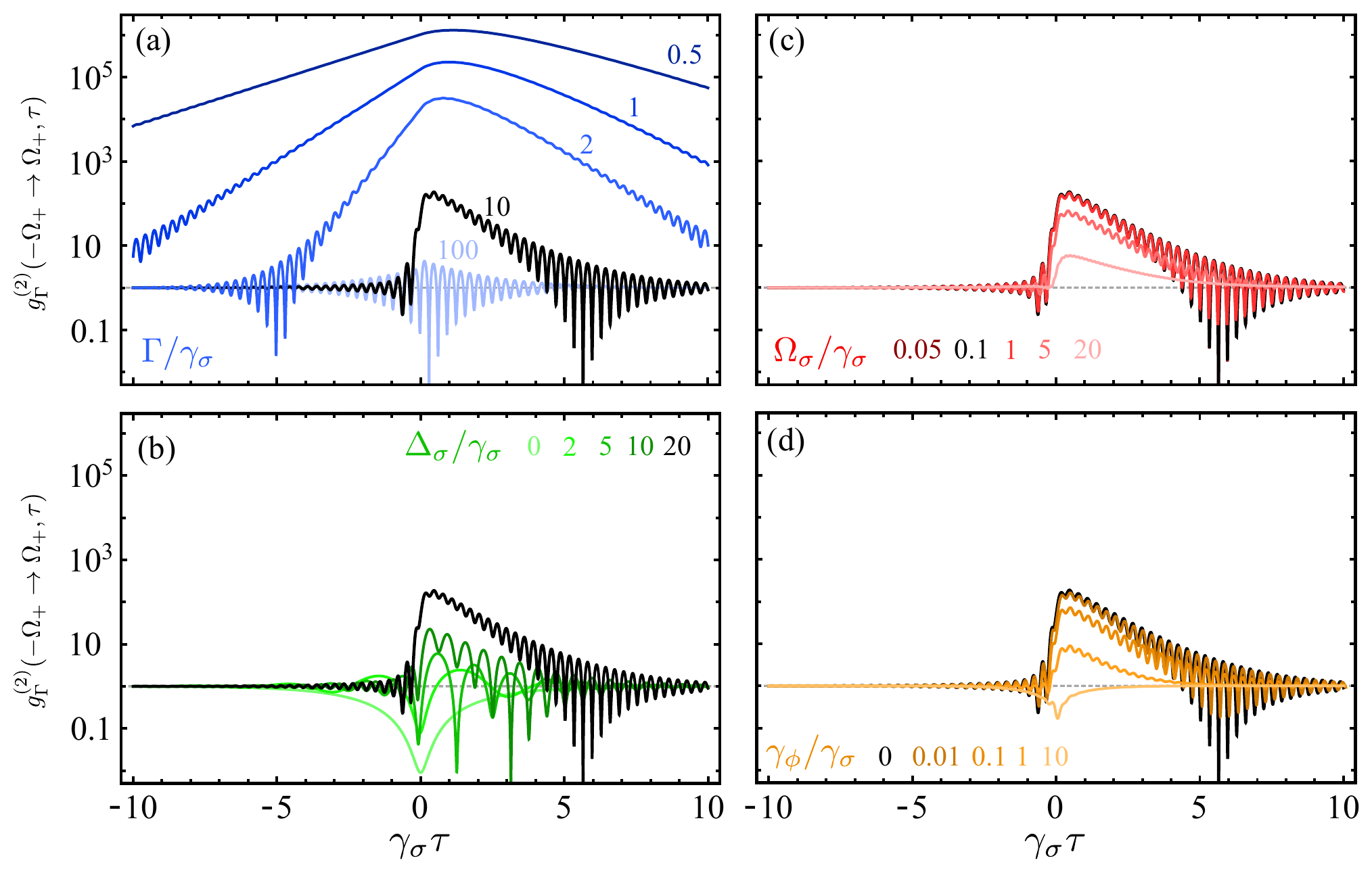}
  \caption{Cross-correlations of the two peaks of detuned resonance
    fluorescence in the cases where, (a)~the filter width is varied,
    (b)~the detuning is varied, (c)~the driving intensity is varied
    or~(d)~the dephasing rate is varied, all other parameters being
    otherwise kept fixed to $\Gamma/\gamma_\sigma=10$, $\Omega=0.1$,
    $\Delta_\sigma/\gamma_\sigma=20$, $\gamma_\phi=0$ and
    $\gamma_\sigma$ setting the unit. The black trace is the same in
    all panels. This figure is discussed in more details in the text
    but is particularly important in connection to
    Fig.~\ref{fig:Tue27Sep203853CEST2022}.}
  \label{fig:Tue27Sep183519CEST2022}
\end{figure}

So far, we have not actually filtered the emission. Instead, we used
homodyning to remove a particular part of the spectrum, namely, the
central scattering peak. We now bring filtering to similarly remove
this central peak, by placing two detectors (or filters) on the side
peaks, although what we actually gain in this way is the possibility
to turn to cross-correlations. This can tell us about the cascaded
emission, i.e., whether one photon comes before the other, which was
the actual prediction from the pioneering papers of the two-photon
character in this case, as opposed to simultaneous emission.  We first
provide the ideal two-photon cascade
cross-correlation~$g^{(2)}(\tau)$, defined as an uncorrelated
(Poisson) stream~($1$) of photons with emission rate~$\gamma_1$, each
photon of which triggers the emission of another photon in another
stream ($2$), in the good time order, i.e., emitted at a later time, with
probability~$p$ and with decay rate~$\gamma_2$, or in the wrong
time-order, i.e., emitted before the triggering photon, wih
probability~$1-p$ and decay rate~$\bar\gamma_2$. In this case, we find
that:
\begin{equation}
  \label{eq:Sun9Oct102241BST2022}
  g^{(2)}(\tau)=1+
  \begin{cases}
    (1-p)\displaystyle{\bar\gamma_2\over\gamma_1}\exp(\bar\gamma_2\tau) & \text{if~$\tau<0$},\\
    p\displaystyle{\gamma_2\over\gamma_1}\exp(-\gamma_2\tau) & \text{if~$\tau>0$}.\\
  \end{cases}
\end{equation}
Note that there is a discontinuity at~$\tau=0$ which prevents to
define $g^{(2)}(0)$, that is however of no concern since this occurs
at a single point (in a practical context, one can take either limit
$\pm\tau\to 0$ or an average). A physical mechanism approaching this
ideal scenario would besides connect smoothly the two domains. In our
case, the exact numerical results for the filtered two-photon
correlations of resonance fluorescence are shown in
Fig.~\ref{fig:Tue27Sep183519CEST2022} for various critical parameters
being varied one at a time.  Taken in turns, we find that:
\begin{itemize}
\item[(a)] With no (or very broad, $\Gamma\gg\gamma_\sigma$)
  filtering, there is no asymmetry and there is a strong antibuching,
  recovering in this case the full correlation of the complete
  spectrum. As the filtering tightens around the peaks (i.e., $\Gamma$
  decreases) an asymmetry develops in the form that is typical of a
  two-photon cascade, cf.~Eq.~(\ref{eq:Sun9Oct102241BST2022}), with a
  transition to bunching that is maximum at positive~$\tau$. The
  correlations also exhibit strong oscillations. For smaller still
  values of~$\Gamma$ (narrow filters), correlations increase but the
  asymmetry reduces. In the limit of $\Gamma\to0$, correlations
  diverge and become~$\tau$ independent (flat).
\item[(b)] As a function of detuning~$\Delta_\sigma$, one goes from
  antibunching at resonance~($\Delta_\sigma=0$) to a growth of the
  asymmetry and increase of the correlations. In this case, the trend
  is consistent and both the asymmetry and correlations increase to
  realize the case of an increasingly better two-photon cascade.
\item[(c)] As a function of driving~$\Omega_\sigma$, one goes from the
  Heitler ($\Omega_\sigma\to0$) to the detuned Mollow (growing an
  incoherent central peak) and ultimately to the resonant Mollow
  triplet~$\Omega_\sigma\gg\gamma_\sigma$ as detuning becomes
  negligible. In this case, the cascade asymmetry remains constant
  while the strength of the correlations decrease and the oscillations
  dampen.
\item[(d)] Finally, as a function of pure dephasing~$\gamma_\phi$, the
  asymmetry also remains constant as the correlations weaken, although
  in this case better retaining the oscillations and also recovering
  some antibunching for very large dephasing, due to spectral overlap
  induced by the line broadening.
\end{itemize}

These observations are informative regarding the general character of
the two-photon emission, which we can summarize as follows: there is a
clear cascade, i.e., one photon comes before the other, as is
consistent with the two-photon scattering picture; this is better
resolved when photons are adequately detected, i.e., there is an
optimum filter width to optimize the effect; the cascade gets better
with increasing detuning, forcing the system into the perturbative
two-photon limit; correlations are stronger for weaker driving, i.e.,
the Heitler regime is more correlated than the Mollow one but
oscillations are also more pronounced; dephasing damages the process,
and even reverts to the opposite regime of single-photon
emission. Also, it should be emphasized once more that these results
are supposedly exact, since they have been obtained with no
approximations from the filtered and photo-detection side once given
the model (which is that from Mollow).  It becomes particularly
interesting, however, to now combine homodyning and cross-correlation
filtering, which we do in the next Section.


\section{Cross-correlations of the homodyned side peaks}

\begin{figure}
    \centering
  \includegraphics[width=\linewidth]{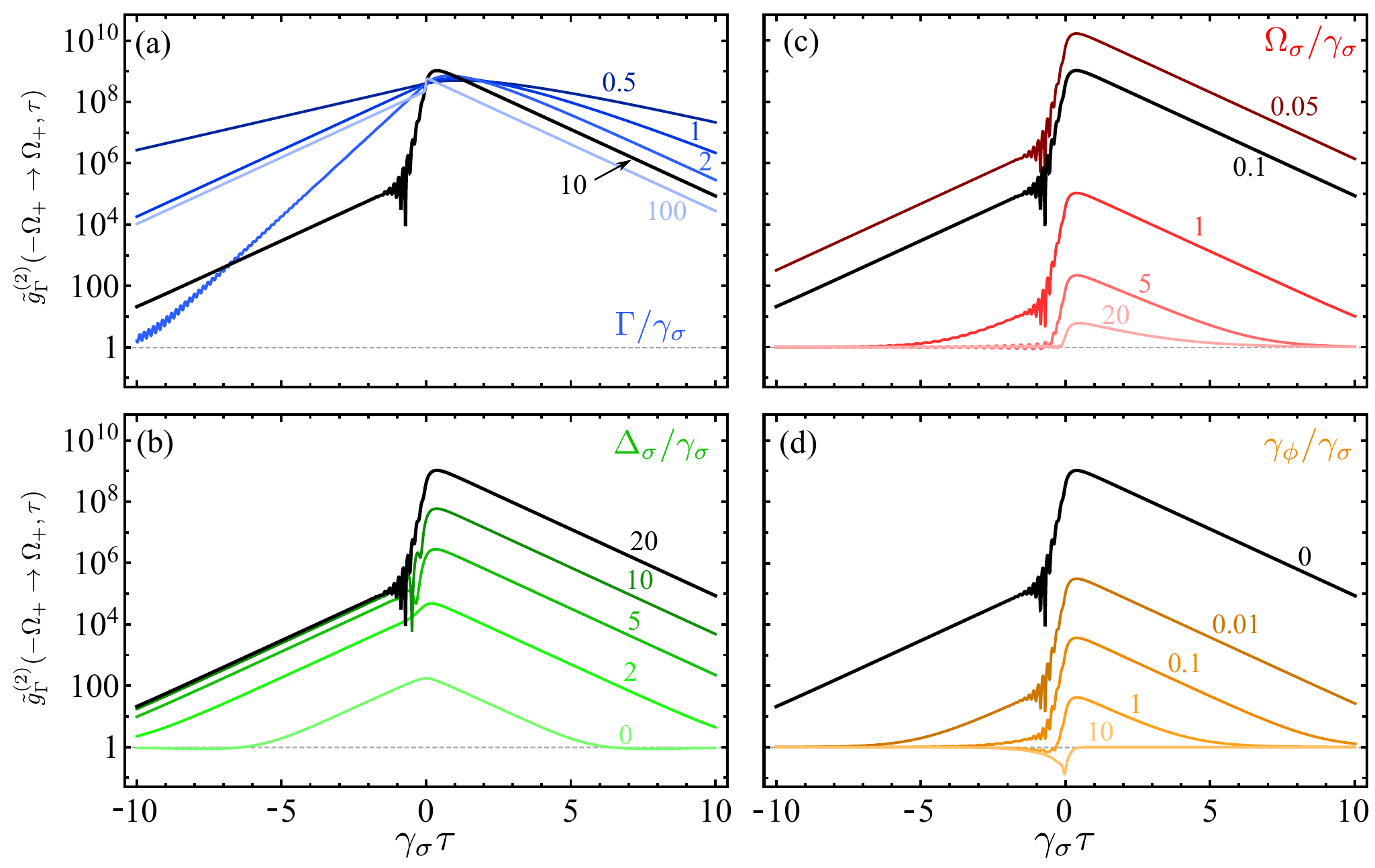}
  \caption{Same as Fig.~\ref{fig:Tue27Sep183519CEST2022} but without
    the coherent peak, i.e., with homodyning~$\mathcal{F}=1$ to
    suppress it. This results in a considerable improvement in the
    two-photon cascade character of the emission.}
  \label{fig:Tue27Sep203853CEST2022}
\end{figure}

We now repeat the same procedure as previously, cross-filtering the
two peaks, but removing the central peak. We do that by homodyning,
since this is the ideal scenario, but this could experimentally be
approached by cross-correlating the output of Masters \emph{et al.}'s
narrowband notch filter. The results are shown in
Fig.~\ref{fig:Tue27Sep203853CEST2022} and are to be contrasted
one-to-one with those of Fig.~\ref{fig:Tue27Sep183519CEST2022}, which
was including the central coherent peak. While one could assume that
cross-correlating the side peaks is tantamount in the first place to
remove the central peak, interference filters have tails and their
suppression is not perfect, therefore some residual physics of
interference leaks through and impact the results. This is why
Fig.~\ref{fig:Tue27Sep194057CEST2022} differs from Masters \emph{et
  al.}'s filter approach: their suppression of the coherent peak in
this way is partial only. The impact turns out to be, maybe
surprisingly, momentous. One need only compare the strong, both
qualitative and quantitative, departures between the non-homodyned and
homodyned versions to appreciate the importance and value of this
modus operandi. First, notice that the scale up to~$\approx 10^6$ in
Fig.~\ref{fig:Tue27Sep183519CEST2022} need be inflated to
over~$10^{10}$ in Fig.~\ref{fig:Tue27Sep203853CEST2022}: correlations
are considerably stronger. Also, the various cases split neatly the
ones from the others, at the exception of filtering, which tends to
now feature more similar magnitudes of correlations, while it retains
an optimum filter width to maximize the asymmetry. Finally,
oscillations are very much, sometimes completely
suppressed. Understanding these oscillations as a way for the system
to provide one-photon emission from a two-photon mechanism by
synchronizing them, this means that, by removing the central peak, we
manage to enter more deeply into a more genuine two-photon
physics. The shapes observed in this case are very close to the ideal
two-photon cascade, Eq.~(\ref{eq:Sun9Oct102241BST2022}), with
identical slopes and thus decay rates~$\gamma_2=\bar\gamma_2$ for the
ordered and out-of-order photons, as expected for a biphoton. The rate
is
$\gamma_2=
\gamma_\sigma-{8\gamma_\sigma\Omega_\sigma^2/(\gamma_\sigma^2+4\Delta_\sigma^2)}\approx\gamma_\sigma$
in the Heitler regime while the fast-rising slope bridging the gap is
given by~$\Gamma$, the filters' width.  For~$\gamma_2=\bar\gamma_2$,
this gap allows from Eq.~(\ref{eq:Sun9Oct102241BST2022}) to estimate
both the ratio~$\gamma_2/\gamma_1$ (absolute gap obtained by the
differences between the $\pm\tau\to 0$ limits) and the probability~$p$
(relative gap obtained by their ratio for high enough correlations),
from which one can see that the time-ordering is very good, with
the~$\omega_\upsilon$ photon arriving with high probabiliy before
the~$\omega_\sigma$ one. It should be noted, however, that two-photon
emission is not fully captured by the two-photon correlations
alone. For instance, the probability that the emission of one photon
successfully heralds the other is not captured by a~$g^{(2)}$
measurement, which is not affected by photon-losses, although the
symmetry of the peaks in luminescence suggests that this also is very
high. Still, a thorough description of the two-photon emission in this
regime is a separate problem. Nevertheless, all these facts together
confirm that an ideal two-photon cascade can be approached in detuned
resonance fluorescence, at large detunings, for filter widths
commensurable with the width of the peaks and when suppressing the
coherent Rayleigh peak.

We conclude with an unexpected result. While the process is more
efficient without dephasing, its correlations are not seriously
affected by it, but the impact of dephasing on the signal itself is
dramatic. Two-photon emission from the side peaks of detuned resonance
fluorescence is very weak in intensity, being second-order in the
driving. Most of the emission originates from the central peak, and
removing it, naturally leaves only little emission. It is interesting
that a tiny amount of pure dephasing changes considerably both the
spectral shape and the amount of incoherent emission:
the perfectly symmetric spectral shape collapses with very small
dephasing, at the same time as the intensity grows considerably. The
symmetric peak at~$\omega_\upsilon$ remains the same (i.e., remains
very small), but a channel is now opened for bare, incoherent emission
from the real transition at~$\omega_\sigma$.  This means that the
two-photon diagrammatic picture breaks down and get substituted by a
more straightforward picture of a classical oscillator driven out of
resonance and emitting at its natural frequency, which is a
first-order process, and therefore of considerably higher intensity
than the two-photon emission. What is maybe more remarkable is that
such first-order processes get cancelled in the coherent picture,
i.e., the interferences leading to one-photon emission contrive to
suppress this channel, that would be expected in a classical
picture. Dephasing plays the role of scrambling this interference by
marring the multiphoton wave interference, thus re-opening a strong
classical channel. We do not think that this feature has been
previously noted in this particular context, although it seems to be
of general validity and is certainly occuring in other systems and
configurations. In our case, it achieves the demonstration that
two-photon emission is a fragile quantum coherent process, of weak
intensity. The loss of the satellite peak on the other side from the
emitter is also more dramatic as~$\Omega_\sigma$ goes to zero, showing
that this is a Heitler, two-photon effect.  Experimentally, it is
likely that some amount of dephasing is present but is counteracted by
some amount of driving. Observing a collapse of the satellite
peak---the more abruptly and with a greater gain of intensity from the
other (emitter) peak, the lower the driving---would be an experimental
demonstration of two-photon physics directly at the level of
photoluminescence.




\section{Summary and Conclusions}

We discussed the two-photon correlations in detuned resonance
fluorescence, whose spectral shape---two-side peaks sitting around a
central Rayleigh scattered peak (at low enough driving)---lends itself
to a natural interpretation in terms of two-photon scattering, as
presented by Cohen--Tannoudji, Reynaud, Dalibard and others.  In
particular, we contrasted our theoretical results to those aimed by
Masters \emph{et al.} of supressing the coherent peak. They achieved
this by filtering it out and otherwise auto-correlating the emission
that passes through, finding a clear transition from antibunching with
strong-oscillations to bunching with reduced---but always
present---oscillations.  In our case, we can completely remove the
coherent peak by perfect destructive interferences with an external
(homodyning) laser. We find essentially the same result as Masters
\emph{et al.} of a transition from perfect antibunching with strong
oscillations, when the coherent peak is present, to bunching when
removing it, although in our case there is also a disappearance of the
oscillations. We then turn to cross-correlations of the peaks, showing
how they bear strong features of a two-photon cascade indeed, that get
considerably improved by removing the coherent peak also in this
case. This shows that homodyning is a valuable additional concept,
rather than supplement, to frequency filtering, and can lead, in the
case of resonance fluorescence, to an almost ideal cascaded two-photon
emission. These considerations are only a small part of the general
picture of multiphoton correlations and emission from resonance
fluorescence, which generalizes to higher photon numbers, involving
other types of scattering processes (prominently, leapfrog
processes~\cite{delvalle13a,gonzaleztudela13a}) and calling for
further investigation of their fundamental character as well as
relevance for technology. The interpretations of the underlying
physics are interesting but intrinsically limited by one's mental
picture(s). The two-photon diagrams in
Fig.~\ref{fig:Sat24Sep165129BST2022} provide a compelling mechanism of
a coherent quantum character, but they do not include the subtle
multiphoton interferences with the Rayleigh peak and the detection
process, which are crucial for a comprehensive description. As befits
quantum mechanics, the best one can do is to compute observables. All
interpretations will be constrained in one way or another, although
useful applications could derive from them. In that regard, we have
provided a formalism to achieve exact results.  It is remarkable that
the simplest treatment of the simplest system, already at the time of
Mollow, yields the best results while more sophisticated techniques,
more suitable for more complex systems, come with some approximations.
We ourselves suspect that the driven two-level system could be the
richest and most fundamental source of multiphoton physics and that
more complex systems could, instead of bringing additional physics,
result in imperfect limits of this ultimate realization. There are
many platforms to implement this physics, from
atoms~\cite{ortizgutierrez19a} to ions~\cite{hoffges97a} and
molecules~\cite{wrigge08a} passing by semiconductors (quantum
dots~\cite{flagg09a}, spin~\cite{vamivakas09a}, NV
centers~\cite{zhou16a}, etc.), superconducting qubits~\cite{baur09a}
and circuits~\cite{astafiev10a}, and still others~\cite{pigeau15a,
  lagoudakis17a, cui21a}. While the interest for basic and fundamental
multiphoton emission originates from the early days of quantum
electrodynamics, recent progress from both material, technological and
theoretical sides makes it a particularly burgeoning field of study.

\ack EdV acknowledges support from the CAM Pricit Plan (Ayudas de
Excelencia del Profesorado Universitario), TUM-IAS Hans Fischer
Fellowship and projects AEI/10.13039/501100011033 (2DEnLight) and
Sinérgico CAM 2020 Y2020/TCS-6545 (NanoQuCo-CM).

\section*{References}

\bibliographystyle{unsrt}
\bibliography{sci,arXiv,Books}

\end{document}